\documentclass[aps,prd,reprint,onecolumn,notitlepage,nofootinbib]{revtex4-1}

\usepackage{hyperref}
\usepackage{graphicx}
\usepackage{multirow}
\usepackage{amsmath}
\usepackage{bm}
\usepackage{slashed}
\usepackage{epsfig}
\usepackage{amsfonts}
\usepackage{color}
\usepackage{epstopdf}
\usepackage{enumitem}
\usepackage{extarrows}
\usepackage{rotating}
\usepackage{autobreak}
\allowdisplaybreaks
\graphicspath{{figures/}{fig/}}

\newcommand{\dt}{\mathrm{t}}

\newcommand{\ud}{\mathrm{d}}
\newcommand{\ui}{\mathrm{i}}

\newcommand{\non}{\nonumber\\}

\begin{document}
\title{Calculation of Feynman loop integration and phase-space integration \\via  auxiliary mass flow}

\author{Xiao Liu$^{1}$}
\email{xiao6@pku.edu.cn}
\author{Yan-Qing Ma$^{1,2,3}$}
\email{yqma@pku.edu.cn}
\author{Wei Tao$^{1}$}
\email{ygtw@pku.edu.cn}
\author{Peng Zhang$^{1}$}
\email{p.zhang@pku.edu.cn}
\affiliation{
$^{1}$School of Physics and State Key Laboratory of Nuclear Physics and
Technology, Peking University, Beijing 100871, China\\
$^{2}$Center for High Energy Physics,
Peking University, Beijing 100871, China\\
$^{3}$Collaborative Innovation Center of Quantum Matter,
Beijing 100871, China
}
\date{\today}

\begin{abstract}
We extend the auxiliary-mass-flow (AMF) method originally developed for Feynman loop integration to calculate integrals involving also phase-space integration. Flow of the auxiliary mass from the boundary ($\infty$) to the physical point ($0^+$) is obtained by numerically solving differential equations with respective to the auxiliary mass. For problems with two or more kinematical invariants, the AMF method can be combined with traditional differential equation method by providing systematical boundary conditions and highly nontrivial self-consistent check. The method is described in detail with a pedagogical example of $e^+e^-\rightarrow \gamma^* \rightarrow t\bar{t}+X$ at NNLO.
We show that the AMF method can systematically and  efficiently calculate  integrals  to high  precision. 
\end{abstract}

\maketitle

\section{Introduction}

With the good performance of the LHC, the particle physics enters an era of precision measurement. To further test the particle physics standard model and to probe new physics, theoretical calculation at high order in the framework of perturbative quantum field theory is needed to match the precision of experimental data. One of the main difficulty for high-order calculation is the phase-space integration. On the one hand, usually there are soft and collinear divergences under integration which makes it impossible to calculate phase-space integration directly using Monte Carlo numerical method. On the other hand, in general it is hard to express results in terms of known analytical special functions.
Significant progresses have been obtained in the past decades.

The mainstream strategy to calculate divergent phase-space integration is to divide integrals into singular part and finite part, so that the first part can be calculated easily (either analytically or numerically) and the second part can be calculated purely numerically using Monte Carlo \cite{GehrmannDeRidder:2005cm,Currie:2013vh,Czakon:2010td,Czakon:2011ve,Boughezal:2011jf,Catani:1996vz,DelDuca:2015zqa,Harris:2001sx,Binoth:2000ps,Borowka:2017idc,Cacciari:2015jma,Catani:2007vq,Boughezal:2015eha,Boughezal:2015dva,Gaunt:2015pea,Caola:2017dug,Magnea:2018hab,Magnea:2018ebr}.
If the process in consideration is sufficient inclusive, one can map phase-space integrals onto corresponding loop integrals by using the reverse unitarity relation~\cite{Anastasiou:2002yz,Anastasiou:2002qz,Anastasiou:2003yy}
\begin{align}\label{eq:invU}
(2\pi) \delta(\mathcal{D}^{\mathrm{c}}_i)
=\frac{\ui}{\mathcal{D}^{\mathrm{c}}_i+\ui0^+}+\frac{-\ui}{\mathcal{D}^{\mathrm{c}}_i-\ui0^+},
\end{align}
where $\mathcal{D}^{\mathrm{c}}_i =
k_i^2-m_i^2$
can be  interpreted either as mass shell condition or as inverse propagator on cut.
In this way, techniques developed for calculating loop integration can be used, like integration-by-parts (IBP) relations~\cite{Chetyrkin:1981qh}, 
differential equations~\cite{Kotikov:1990kg,Gehrmann:1999as},  dimensional recurrence relations~\cite{Tarasov:1996br,Lee:2009dh,
Lee:2017ftw}, and also methods developed by introducing auxiliary mass (AM)~\cite{Liu:2017jxz,Liu:2018dmc,Zhang:2018mlo,Wang:2019mnn,Guan:2019bcx,Yang:2020msy,Bronnum-Hansen:2020mzk}. Further more, loop integration and phase-space integration can be dealt with as a whole because they are not significantly different from each other for these techniques.

\begin{figure}[htbp]
	\begin{center}
		\includegraphics[width=0.4\linewidth]{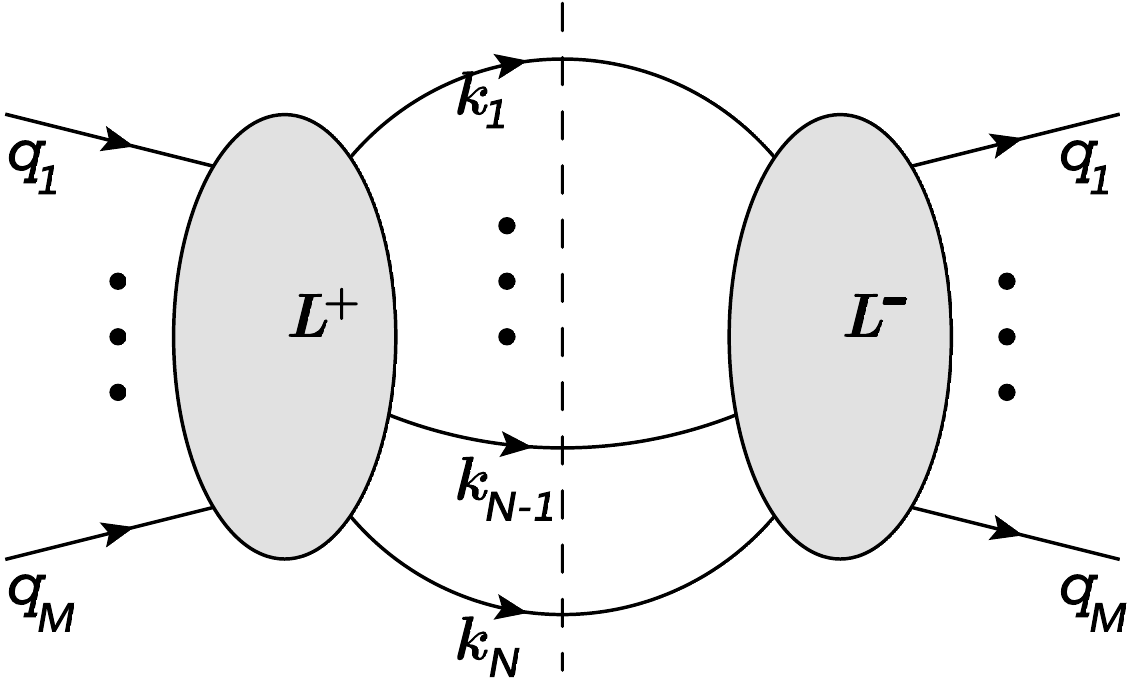}
		\caption{\label{fig:process} A schematic diagram for a process with  $L=L^++L^-$ loops, $M$ unintegrated external legs, and $N$ cut legs.}
	\end{center}
\end{figure}

To be definite, a schematic cut diagram for a general process is shown in Fig.\ref{fig:process}, where $L^+$ is the number of loop momenta (denoting as $\{l_i^+\}$) on the l.h.s. of the cut, $L^-$ is the number of loop momenta (denoting as $\{l_i^-\}$) on the r.h.s. of the cut, $L=L^++L^-$ is the total number of loop momenta, $M$ is the  number of external momenta (denoting as $\{q_i\}$) which contains not only initial state  external momenta but also fixed and unintegrated final state momenta, and $N$ is the   number  of  cut  momenta (denoting as $\{k_i\}$) which are on mass shell  with $m_i$ being corresponding  particle mass. $L^+\geq0$, $L^-\geq0$, $M\geq1$ and $N\geq1$ are understandable. We denote $Q$ as the total cut  momentum  which satisfies $Q=\sum_{i=1}^M q_i=\sum_{i=1}^N k_i$ if $\{q_i\}$ are labeled to flow into the diagram and  $\{k_i\}$ flow out of the diagram.
A complete set of kinematical invariants after performing loop integration and phase-space integration  is denoted as $\vec{s}$ with $Q^2$ as a special component.

A general phase-space and loop integration with AM to be studied in this work is
\begin{align}\label{eq:Ps}
{F}(\vec{\nu};\vec s, \vec \eta)\equiv\int \text{dPS}_N \prod_{\alpha}\frac{1}{(\mathcal{D}^{\dt}_{\alpha}+\eta^{\dt}_\alpha)^{\nu^{\dt}_{\alpha}}}
\int \prod_{i=1}^{L^+} \frac{\ud^Dl_i^+}{(2\pi)^D}\prod_{\beta}\frac{1}{(\mathcal{D}^+_{\beta}+\ui\eta^+_\beta)^{\nu^+_{\beta}}}
\int \prod_{j=1}^{L^-}\frac{\ud^Dl_j^-}{(2\pi)^D}\prod_{\gamma}\frac{1}{(\mathcal{D}^-_{\gamma}-\ui\eta^-_\gamma)^{\nu^-_{\gamma}}}
(l_i^+\cdot l_j^-)^{-\nu^{\pm}_{ij}} ,
\end{align}
where the non-integer spacetime dimension $D=4-2\epsilon$ is introduced to regularize all possible divergences, $\mathcal{D}^{\dt}_{\alpha}$ are inverse of tree propagators which do not depend on loop momenta,  $\mathcal{D}^+_{\alpha}$ are inverse of loop propagators on the l.h.s. of the cut which do not depend on loop momenta on the r.h.s. of the cut,  $\mathcal{D}^-_{\alpha}$ are inverse of loop propagators on the r.h.s. of the cut which do not depend on loop momenta on the l.h.s. of the cut,  the vector $\vec{\nu}\equiv({\nu}^{\dt}_1,{\nu}^{\dt}_2,\cdots,{\nu}^+_1,{\nu}^+_2,\cdots,{\nu}^-_1,{\nu}^-_2,\cdots,\nu^{\pm}_{11},\nu^{\pm}_{12},\nu^{\pm}_{21},\cdots)$ with $\nu_{ij}^\pm\leq0$, the vector $\vec{\eta}\equiv({\eta}^{\dt}_1,{\eta}^{\dt}_2,\cdots,{\eta}^+_1,{\eta}^+_2,\cdots,{\eta}^-_1,{\eta}^-_2,\cdots)$ denotes the introduced AM terms, and $\text{ dPS}_N$ is the measure of $N$-particle-cut phase-space integration. For total cross section, we have
\begin{align}\label{eq:defPS}
\text{ dPS}_N\equiv(2\pi)^{D}\delta^D(Q-\sum_{i=1}^N k_i)\prod_{i=1}^{N}\frac{\ud^Dk_i}{(2\pi)^{D}}(2\pi)\delta(\mathcal{D}^{\mathrm{c}}_i)\Theta\!(k_{i}^{0}-m_i),
\end{align}
where $\Theta$ is the Heaviside function. \footnote{Note that we use $\Theta\!(k_{i}^{0}-m_i)$ instead of usual $\Theta\!(k_{i}^{0})$ here. Although they are equivalent in the definition of Eq.~\eqref{eq:defPS}, the advantage of $\Theta\!(k_{i}^{0}-m_i)$ is that its derivative (being Dirac delta function) can be safely set to zero in dimensional regularization. It is guaranteed by the fact that Dirac delta function restricts all space components of $k_i$ to be at the origin, where is well regularized by dimensional regularization. Therefore, our choice is convenient to use inverse unitarity. } Differential cross sections can be obtained by introducing constraints into $\text{ dPS}_N$.

The corresponding physical integral can be obtained from the above modified integral by taking all AMs to zero,
\begin{align}
{F}(\vec{\nu};\vec s,0)\equiv \lim_{\vec\eta\to0^+}{F}(\vec{\nu};\vec s,\vec \eta).
\end{align}
It is understandable to take $\eta^+_\alpha$ and $\eta^-_\alpha$ to zero from the positive side of their real parts, because this is exactly the rule of Feynman prescription for Feynman propagators which guarantees the correct discontinuity of Feynman propagators. While for $\eta^{\dt}_\alpha$, we can take them to zero from any direction as far as all tree propagators are either positive-definite or negative-definite. Our choice is to take all $\eta^{\dt}_\alpha$ to zero from the positive side so that tree propagators on the l.h.s. of the cut can be combined with the same propagators on the r.h.s. of the cut. \footnote{Positive (or negative) definiteness of tree propagators is alway satisfied in the narrow-width approximation, where particle production and decay are factorized and can be calculated separately. Otherwise, we should distinguish $\eta^{\dt}_\alpha$ on the two sides of the cut and take it to $\ui 0^+$ or $-\ui 0^+$ respectively.}

For our purpose, we can choose components of $\vec \eta$ to be  either fully related to each others or completely independent. An extreme is to choose all components of $\vec \eta$ to be the same, and an opposite extreme is to choose a strong ordering for all components of $\vec \eta$. Although all these choices are workable, to be definite in this work we assume that components of $\vec \eta$ can only be either $0^+$ or $\eta$. In this way, $F$ only depends on one AM $\eta$, and we denote it as ${F}(\vec{\nu}; \vec s, \eta)$ in the rest of this paper.

In this work, we study the  calculation of physical ${F}(\vec{\nu};\vec s,0)$ based on the auxiliary-mass-flow (AMF) method originally proposed in Ref.~\cite{Liu:2017jxz} for pure loop integration, where flow of $\eta$ from $\infty$ to $0^+$ is obtained by solving differential equations w.r.t. $\eta$. We will see that this method is not only systematical and efficient, but also can give high-precision result.
The rest of the paper is organized as follows.
In  Sec.\ref{sec:pssrde}, we describe the  general strategy to  calculate  integrals involving both loop integration and phase-space integration.
In Sec.\ref{sec:psexamp}, the method is explained in detail by a pedagogical example $e^+e^-\rightarrow \gamma^* \rightarrow t\bar{t}+X$ at NNLO. We also verify the correctness of our calculation by various methods.
Finally, a summary is given in Sec.\ref{sec:pssum}.
The calculation of basal phase-space integration without denominator in the integrand is given in Appendix~\ref{sec:pscut}.

\section{Auxiliary mass expansion and flow}\label{sec:pssrde}

The advantage of introducing $\eta$ is that, by taking $\eta\to \infty$, ${F}(\vec{\nu};\vec s, \eta)$ can be reduced to linear combinations of simpler integrals. As  scalar products  among external momenta and cut momenta  are finite, we have the auxiliary mass expansion (AME) for tree propagators
\begin{align}
\frac{1}{\mathcal{D}^{\dt}_\alpha+\eta}&\xlongequal{\eta\to\infty}
\frac{1}{\eta}\sum_{j=0}^{+\infty}\left(\frac{-\mathcal{D}^{\dt}_\alpha}{\eta}\right)^j,\\
\frac{1}{\mathcal{D}^{\dt}_\alpha}&\xlongequal{\eta\to\infty}
\frac{1}{\mathcal{D}^{\dt}_\alpha},
\end{align}
which removes a tree propagator from the denominator if $\eta$ has been introduced to it. Because loop momenta can be any large value, one cannot naively expand loop propagators in the same way as tree propagators. However, the standard rules of  large-mass expansion \cite{Smirnov:1990rz,Smirnov:1994tg} imply that, as $\eta\to\infty$, linear combinations of loop momenta can be either at the order of ${|\eta|}^{1/2}$ or much smaller than it. Therefore one can do the following AME,
\begin{align}\label{eq:expeta}
\frac{1}{\mathcal{D}^+_\alpha+\ui \eta}&\xlongequal{\eta\to\infty}
\frac{1}{\widetilde{\mathcal{D}}^+_{\alpha}+\ui\eta}\sum_{j=0}^{+\infty}\left(\frac{-K_\alpha}{\widetilde{\mathcal{D}}^+_{\alpha}+\ui\eta}\right)^j,
\end{align}
\begin{align}   
\frac{1}{\mathcal{D}^+_\alpha+\ui 0^+}\xlongequal{\eta\to\infty}
\begin{cases}
\frac{1}{\widetilde{\mathcal{D}}^+_{\alpha}+\ui0^+}\sum_{j=0}^{+\infty}\left(\frac{-K_\alpha}{\widetilde{\mathcal{D}}^+_{\alpha}+\ui0^+}\right)^j  &  \text{if $\widetilde{\mathcal{D}}^+_{\alpha}\neq0$,} \\
\frac{1}{{\mathcal{D}}^+_{\alpha}+\ui0^+}
&  \text{if $\widetilde{\mathcal{D}}^+_{\alpha}=0$,}
\end{cases}               
\end{align}
where we decompose $\mathcal{D}^+_\alpha=\widetilde{\mathcal{D}}^+_{\alpha}+K_\alpha$ with $\widetilde{\mathcal{D}}^+_{\alpha}$ including only the part at the order of ${|\eta|}$. Similarly we can do the expansion for loop propagators on the r.h.s. of the cut. The AME of loop propagators either removes some propagators from the denominator (if $\eta$ presents in the propagator and $\widetilde{\mathcal{D}}^+_{\alpha}=0$) or decouples some loop momenta at the order of ${|\eta|}^{1/2}$ from kinematical invariants. The later effect results in some single-scale vacuum integrals multiplied by integrals with less number of loop momenta. 

We find that, as $\eta\to \infty$, ${F}(\vec{\nu};\vec s, \eta)$ is simplified to linear combination of integrals with less inverse propagators in the denominator (maybe multiplied by single-scale vacuum integrals). If the simplified integrals still have inverse propagators in the denominator (except propagators in single-scale vacuum integrals), we can again introduce new AM $\eta$ and take $\eta\to \infty$. Eventually, ${F}(\vec{\nu};\vec s, \eta)$ is translated to the following form
\begin{align}
{F}(\vec{\nu}; \vec s, \eta)\longrightarrow\sum c\times F^{\mathrm{cut}} \times  F^{\mathrm{bub}},
\end{align}
where $c$ are rational functions of $\vec s$ and $\eta$, $F^{\mathrm{bub}}$ include only single-scale vacuum bubble integrals, and  $F^{\mathrm{cut}}$ denote  basal  phase-space integrations with integrand being polynomials of scalar products between   cut momenta. $F^{\mathrm{bub}}$ have been well studied up to five-loop order~\cite{Davydychev:1992mt,Broadhurst:1998rz,Kniehl:2017ikj,Schroder:2005va,Luthe:2015ngq,Luthe:2017ttc}. $F^{\mathrm{cut}}$ can be easily dealt with because the only nontrivial information are the cut propagators, which will be explicit studied in Appendix.\ref{sec:pscut}. With these information in hand, the next question is how to obtain physical integrals.

It was shown in Ref.~\cite{Smirnov:2010hn} that, for any given problem,  Feynman loop integrals form an finite-dimensional vector space, with basis of which called master integrals (MIs). The step to express all loop integrals as linear combinations of MIs are called reduction. With reverse unitarity relation in Eq.\eqref{eq:invU}, one can map phase-space integrations onto corresponding loop integrations. Therefore, integrals over phase-space and loop momenta defined in Eq.~\eqref{eq:Ps} can also be reduced to corresponding MIs.

Reduction of a general integral to MIs can be traditionally achieved by using IBP relations based on Laporta's algorithm~\cite{Laporta:2001dd,Anastasiou:2004vj,Smirnov:2019qkx,
Maierhofer:2018gpa,vonManteuffel:2012np,Lee:2013mka}. Alternatively, one can achieve IBP reduction using finite-field
interpolation~\cite{vonManteuffel:2014ixa,Peraro:2016wsq,Klappert:2019emp,Klappert:2020aqs,Klappert:2020nbg,Peraro:2019svx}, module intersection~\cite{Boehm:2018fpv}, intersection theory~\cite{Mastrolia:2018uzb}, or AME \cite{Liu:2018dmc}.  \footnote{ IBP reduction can be achieved by AME for the specific ${F}(\vec{\nu}; \vec s, \eta)$ with $\eta$ introduced to {\it{all}} inverse propagators (not for cut propagators), which is a generalization of the method for pure loop integration introduced in Ref. \cite{Liu:2018dmc}. With this strategy, coefficients of the expansion  are polynomials of kinematical invariants.} In any case, the search algorithm proposed in Refs. \cite{Liu:2018dmc,Guan:2019bcx} can significantly improve the efficiency of reduction, which makes the reduction of very complicated problems a possibility.
Reduction can not only express all integrals in terms of MIs, it can also set up differential equations (DEs) among MIs~\cite{Kotikov:1990kg,Bern:1992em,Remiddi:1997ny,Gehrmann:1999as,Henn:2013pwa,Lee:2014ioa}. Especially, DEs w.r.t. the AM $\eta$ are given by
\begin{align}
\frac{\partial}{\partial \eta} \vec{J}(\vec s, \eta) = M(\vec s, \eta) \vec{J}(\vec s, \eta),
\end{align}
where $\vec{J}(\vec s, \eta)\equiv \left\{ {F}(\vec{\nu}', \vec s, \eta), {F}(\vec{\nu}'', \vec s, \eta), \cdots\right\}$ is a complete set of MIs and $M(\vec s, \eta)$ is the coefficient matrix as rational function of $\vec s$ and $\eta$.
Boundary condition of the DEs can be chosen at $\eta\to\infty$, which can be easily obtained by the AME discussed above. By solving the above DEs (usually numerically) one can realize the flow of $\eta$ from $\infty$ to $0^+$. In this way, we get a general method to calculate physical MIs $\vec{J}(\vec s, 0)$ with any fixed $\vec s$.

Furthermore, in the case of more than one kinematical invariant, the AMF method can also be combined with DEs w.r.t. $\vec{s}$ to obtain MIs at different values of $\vec{s}$.

\section{Examples: master integrals for $e^+e^-\rightarrow \gamma^* \rightarrow t\bar{t}+X$ at NNLO}\label{sec:psexamp}

As a simple but nontrivial example, we calculate MIs encountered in the NNLO correction for $t\bar{t}$ production in $e^+ e^-$ collision mediated by a virtual photon  to demonstrate the validity of AMF method. For the purpose of total cross section, there are only two kinematical invariants ($Q^2$ and $m_t^2$) besides $\eta$. Thus we can introduce dimensionless integrals
\begin{align}
\hat{F}(\vec{\nu};x,y)\equiv s^{N-\frac{N-1+L}{2}D+\nu} {F}(\vec{\nu};\vec{s},\eta),
\end{align}
where $s=Q^2$, $x=\frac{4m_t^2}{s}$, $y=\frac{\eta}{s}$ and $\nu$ is the summation of all components of $\vec \nu$. 
Because the problem is simple, we make the following unoptimized scheme choice:
\begin{align}
{\eta}^t_1&={\eta}^{\dt}_2=\cdots, \nonumber\\
{\eta}^+_1&={\eta}^+_2=\cdots=
{\eta}^-_1={\eta}^-_2=\cdots,
\end{align}
with $\eta_1^{\dt}\ll \eta_1^+$. More precisely, if $\{\mathcal{D}^-_{\alpha}\}$ or $\{\mathcal{D}^+_{\alpha}\}$ depend on $\vec s$, we choose $\eta^{\dt}_1= 0^+$ and $ \eta^+_1= \eta$; Otherwise, we choose $\eta^{\dt}_1= \eta$ (introduction of $\eta_1^+$ is unnecessary in this case).
The publicly available systematic package FIRE6~\cite{Smirnov:2019qkx} is sufficient to do all needed reductions in this simple exercise.

There are many subprocesses at the partonic level. We use $\mathrm{V}^{L^+}\mathrm{R}^{N-1}\mathrm{V}^{L^-}$ to distinguish processes with different number of independent loop momenta and cut momenta. The presence of $N-1$ in stead of $N$ is a result of momentum conservation which reduces one independent cut momentum.  For completeness, we first provide MIs at NLO in Sec.\ref{sec:RRR}. Calculation of MIs for 4-particle cuts at NNLO (RRR) will be presented in Sec.\ref{sec:RRRR}. Calculation of MIs for 3-particle cuts at NNLO (VRR) will be presented in Sec.\ref{sec:VRRR}. And calculation of MIs for 2-particle cuts at NNLO (VVR or VRV) will be presented in Sec.\ref{sec:VVRR}. Verification of these results will be given in Sec.\ref{sec:check}. We note that all these results have already been calculated in literature using other methods (see, e.g., Refs.~\cite{Bernreuther:2011jt,Bernreuther:2013uma,Dekkers:2014hna,Magerya:2019cvz}) although they are not fully publicly available. We provide our results as ancillary file in the arXiv version.

\subsection{RR and VR}\label{sec:RRR}
For RR, i.e., $\gamma^*\to t\bar{t}g$  process,  inverse propagators  can be  written as
\begin{align}
&\mathcal{D}^{\mathrm{c}}_1=k_1^2-m_t^2,
\mathcal{D}^{\mathrm{c}}_2=k_2^2-m_t^2,
\mathcal{D}^{\mathrm{c}}_3=(Q-k_1-k_2)^2;\non
&\mathcal{D}^\dt_1=(Q-k_1)^2-m_t^2,
\mathcal{D}^\dt_2=(Q-k_2)^2-m_t^2.
\end{align}
The obtained  2  physical  MIs are $F^\mathrm{cut}_{2,3,1}$ and $F^\mathrm{cut}_{2,3,2}$, which are calculated in Appendix.\ref{sec:pscut}.

For VR, inverse propagators  can be  written as
\begin{align}
&\mathcal{D}^{\mathrm{c}}_1=k_1^2-m_t^2,
\mathcal{D}^{\mathrm{c}}_2=(Q-k_1)^2-m_t^2;\non
&\mathcal{D}^+_1=(k_1+l^+_1)^2-m_t^2,
\mathcal{D}^+_2=(Q-k_1-l^+_1)^2-m_t^2,
\mathcal{D}^+_3={l^+_1}^2.
\end{align}
We obtain 1 family \footnote{A family of integrals are characterized by the kinds of inverse propagators presented in the denominator of the corresponding integrand. As usual, we use the corner integral, which has no inverse propagator in the numerator and has the maximal kinds of inverse propagators in the denominator with power of each inverse propagators being unit, to represent the family.} with $\vec \nu$ of characterized MIs being $\{1,1,0\}$. Physical   MIs are given by
\begin{align}
&\int\text{dPS}_2\int\frac{\ud^Dl^+_1}{(2\pi)^D}\frac{1}{\mathcal{D}^+_1+\ui0^+}=
\Phi_2\int\frac{\ud^Dl^+_1}{(2\pi)^D}\frac{1}{{l^+_1}^2-m_t^2+\ui0^+},\,\nonumber\\
&\int\text{dPS}_2\int\frac{\ud^Dl^+_1}{(2\pi)^D}\frac{1}{(\mathcal{D}^+_1+\ui0^+)(\mathcal{D}^+_2+\ui0^+)}=
\Phi_2\int\frac{\ud^Dl^+_1}{(2\pi)^D}\frac{1}{({l^+_1}^2-m_t^2+\ui0^+)((l^+_1-Q)^2-m_t^2+\ui0^+)},
\end{align}
where $\Phi_2=\int\text{dPS}_2$ is defined in  Eq.\eqref{eq:basmif2} and the remaining 1-loop integrals are easy to calculate analytically.

\subsection{RRR}\label{sec:RRRR}
\begin{figure}[htb!]
\begin{center}
\includegraphics[width=0.8\textwidth]{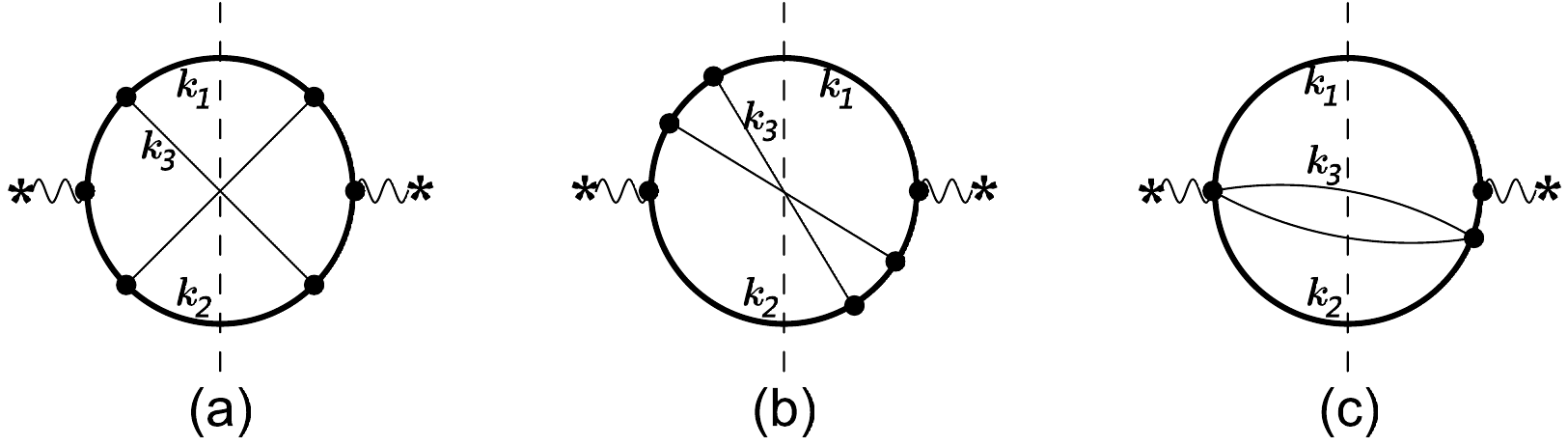}
\caption{\label{fig:RRRR} Representative Feynman diagrams   for $\gamma^*\rightarrow t\bar{t}gg$ or $t\bar{t} q\bar{q}$ process, where (a) and (b) define the two most complicated family and (c) defines a sub-family of (b). Here thick curves represent top quark, thin curves represent massless particle, and vertical dashed lines represent final state cut.}
\end{center}
\end{figure}

For   RRR, i.e., $\gamma^*\to t\bar{t}gg$ or $t\bar{t} q\bar{q}$  processes,  inverse propagators can be written as
\begin{align}
&\mathcal{D}^{\mathrm{c}}_1=k_1^2-m_t^2,
\mathcal{D}^{\mathrm{c}}_2={k_2}^2-m_t^2,
\mathcal{D}^{\mathrm{c}}_3=k_3^2,
\mathcal{D}^{\mathrm{c}}_4=(Q-k_1-k_2-k_3)^2;\,\nonumber\\
&\mathcal{D}^{\mathrm{t}}_1=(Q-k_1-k_2)^2,
\mathcal{D}^{\mathrm{t}}_2=(k_1+k_3)^2-m_t^2,
\mathcal{D}^{\mathrm{t}}_3=(Q-k_2-k_3)^2-m_t^2,
\mathcal{D}^{\mathrm{t}}_4=(Q-k_2)^2-m_t^2,\,\nonumber\\
&\mathcal{D}^{\mathrm{t}}_5=(k_2+k_3)^2-m_t^2,
\mathcal{D}^{\mathrm{t}}_6=(Q-k_1-k_3)^2-m_t^2,
\mathcal{D}^{\mathrm{t}}_7=(Q-k_1)^2-m_t^2,
\end{align}
where  $Q^2=s$. Then phase-space integrals can be expressed as
\begin{align}
\hat{F}(\vec{\nu};x,y)= s^{4-\frac{3}{2}D+\sum \nu^{\mathrm{t}}_{\alpha}} \int\text{dPS}_4\prod_{\alpha=1}^{7}\frac{1}{(\mathcal{D}^{\mathrm{t}}_\alpha+\eta)^{\nu^{\mathrm{t}}_{\alpha}}}.
\end{align}
By using FIRE6, we find that there are 37 MIs for finite $\eta$ and the number is reduced to 15 as $\eta$ vanishing.  \footnote{Note that the number of MIs in this work may not be the minimal value. It does not matter as far as it is finite.}
These MIs can be classified into 2 families with  $\vec \nu$ of characterized MIs being
\begin{align}
\{0,1,1,0,1,1,0\} ~~~\text{and}~~~\{0,1,0,1,1,0,1\},
\end{align}
which can also characterized by Feynman diagrams Fig.\ref{fig:RRRR} (a) and (b), respectively.
Using our method, we can calculate all physical MIs with any fixed $x$. The result of the most complicated MI with, e.g., $x=1/2$ is
\begin{align}
\begin{autobreak}
\MoveEqLeft
\hat F( \{-1, 1, 0, 1, 1, 0, 1\};\frac{1}{2},0)=\hspace{10cm}
4.54087957883195468901389004370\times10^{-7} 
+0.0000105911293014536670979999400899 \epsilon
  + 0.000124406630344529071923178953167 \epsilon^2
  + 0.000983063887963543479220271851806 \epsilon^3 
 + 0.00589205324016960475844728350032 \epsilon^4 
 + 0.0286458793127880349046701435743 \epsilon^5
  + 0.118038608602644851031978155328 \epsilon^6 
 + 0.425508191586298756000765241113 \epsilon^7
  + 1.37520023939856884048640232783 \epsilon^8 +\cdots,
\end{autobreak}
\end{align}
where we truncate the expansion to order $\epsilon^8$ with about 30-digit precision for each coefficient. 
 Physical MIs with other values of $x$ can be calculated similarly.


In the following, let us take a sub-family $\{0,0,0,0,0,0,1\}$ shown in Fig.\ref{fig:RRRR} (c) as an example to illustrate the calculation procedure of MIs.
For brevity, we define MIs for this family as:
\begin{align}
\hat F(\{\nu^{\dt}_1,\nu^{\dt}_2,\nu^{\dt}_7\};x,y)= s^{4-\frac{3}{2}D+\nu^{\dt}_1+\nu^{\dt}_2+\nu^{\dt}_7} \int\text{dPS}_4\frac{(\mathcal{D}^{\mathrm{t}}_1+\eta)^{-\nu^{\dt}_1}(\mathcal{D}^{\mathrm{t}}_2+\eta)^{-\nu^{\dt}_2}}{(\mathcal{D}^{\mathrm{t}}_7+\eta)^{\nu^{\dt}_7}},
\end{align}
with $\nu^{\dt}_1, \nu^{\dt}_2\leq 0$ and $\nu^{\dt}_7\geq0$. This family contains 5 MIs for finite $\eta$ (or $y$)
\begin{align}
\left\{\hat F(\{0,0,0\};x,y),\hat F(\{-1,0,0\};x,y),\hat F(\{0,-1,0\};x,y),\hat F(\{0,0,1\};x,y),\hat F(\{-1,0,1\};x,y)\right\},
\end{align}
and 4 MIs as $\eta\to0^+$
\begin{align}\label{eq:phyMI}
\left\{\hat F(\{0,0,0\};x,0),\hat F(\{-1,0,0\};x,0),\hat F(\{0,-1,0\};x,0),\hat F(\{0,0,1\};x,0)\right\}.
\end{align}
It is clear that the first three MIs of the two sets are  just linear combinations of $F^{\mathrm{cut}}$  that have been calculated in Appendix.\ref{sec:pscut}.

To calculate the last physical MI in Eq.~\eqref{eq:phyMI}, we set up DEs for corresponding MIs with $x=1/2$ and finite $\eta$, which gives
\begin{align}\label{eq:dey}
\begin{split}
&\frac{\partial}{\partial{y}}
\left(
\renewcommand\arraystretch{2.5}
\setlength{\arraycolsep}{1pt}
\begin{array}{c}
\hat F(\{0,0,1\};\frac{1}{2},y)\\
\hat F(\{-1,0,1\};\frac{1}{2},y)
\end{array}
\right)
=
\left(
\renewcommand\arraystretch{2.5}
\setlength{\arraycolsep}{1pt}
\begin{array}{cc}
\frac{ \binom{3 - 4 \epsilon + 14y - 12\epsilon y -8 y^2 }{+52 \epsilon y^2 -
 16 y^3 + 48 \epsilon y^3}}{y (1- 8 y) (1 + 4 y +  2 y^2)}&
\frac{  12 (-1 + \epsilon)}{y (1- 8 y)}   \\
\frac{-4(-1 + \epsilon)}{ 1 - 8 y }&
\frac{  2 \binom{-7 + 6\epsilon - 39 y +46\epsilon y}{  -
      24 y^2 + 32 \epsilon y^2}}{(1- 8 y) (1 +
      4y + 2 y^2) }
\end{array}
\right)
\left(
\renewcommand\arraystretch{2.5}
\setlength{\arraycolsep}{1pt}
\begin{array}{c}
\hat F(\{0,0,1\};\frac{1}{2},y)\\
\hat F(\{-1,0,1\};\frac{1}{2},y)
\end{array}
\right)+\\
&\left(
\renewcommand\arraystretch{2.5}
\setlength{\arraycolsep}{1pt}
\begin{array}{ccc}
\frac{-2 \binom{15- 16\epsilon + 53 y -46 \epsilon y}{  -
      124 y^2 + 132 \epsilon y^2}}{y (1 -
      8 y) (1 + 4y + 2 y^2)}&
\frac{ - 48 (-1 + \epsilon) (1 - y)}{y (1 - 8 y) (1 +
      4 y + 2 y^2) }&
\frac{  -144 (-1 + \epsilon) (1 - y)}{y (1 - 8 y) (1 +
      4 y + 2 y^2) }  \\
\frac{- 2 \binom{17 - 18 \epsilon + 85y }{ -78\epsilon y -
     88 y^2 + 96 \epsilon y^2}}{(1- 8 y) (1 +
     4 y + 2 y^2)   }&
\frac{-56(-1 + \epsilon)}{(1 - 8 y) (1 + 4 y +
     2 y^2)  }&
\frac{ -8 (-1 + \epsilon) (23- 16 y)}{(1 - 8 y) (1 +
     4y + 2 y^2)  }
\end{array}
\right)
\left(
\renewcommand\arraystretch{1.5}
\setlength{\arraycolsep}{1pt}
\begin{array}{c}
\hat F(\{0,0,0\};\frac{1}{2},y)\\
\hat F(\{-1,0,0\};\frac{1}{2},y)\\
\hat F(\{0,-1,0\};\frac{1}{2},y)
\end{array}
\right).
\end{split}
\end{align}
The boundary condition of these DEs is given by
\begin{align}
\begin{split}
\hat F(\{0,0,1\};\frac{1}{2},y)&\xlongequal{\eta\to\infty}s^{5-\frac{3}{2}D}\left(\frac{ 1}{\eta}\int\text{dPS}_4-\frac{1}{\eta^2}  \int\text{dPS}_4\mathcal{D}^{\mathrm{t}}_7\right)_{x=1/2},\\
\hat F(\{-1,0,1\};\frac{1}{2},y)&\xlongequal{\eta\to\infty}s^{4-\frac{3}{2}D}\left( \int\text{dPS}_4-\frac{1}{\eta} \int\text{dPS}_4(\mathcal{D}^{\mathrm{t}}_1-\mathcal{D}^{\mathrm{t}}_7)\right)_{x=1/2}.
\end{split}
\end{align}
By solving the DEs~\eqref{eq:dey} with boundary condition we obtain, e.g.,
\begin{normalsize}
\begin{align}\label{eq:check1}
\begin{autobreak}
\MoveEqLeft
\hat F(\{0,0,1\};\frac{1}{2},0)=
1.08703304446867684362962983900\times10^{-8} 
+ 2.35523252532576916745398951290\times10^{-7} \epsilon 
 + 2.54552584682491818967669600707\times10^{-6} \epsilon^2
  + 0.0000183017607702920607720306203025 \epsilon^3
  + 0.0000985014117985155673601315129686 \epsilon^4
  + 0.000423451568312682839128853965245 \epsilon^5
  + 0.00151532069472999876700566927592 \epsilon^6 
 + 0.00464561935910925265889327482252 \epsilon^7 
 + 0.0124659262732530489584879132787 \epsilon^8  +\cdots.
\end{autobreak}
\end{align}
\end{normalsize}

Finally, the DE w.r.t. $x$ is given by
\begin{align}
\begin{split}
\frac{\partial}{\partial{x}}
\hat F(\{0,0,1\};x,0)
=&
\frac{-2 + 3 x + 4 \epsilon - 6 x \epsilon}{2 (-1 + x) x}
\hat F(\{0,0,1\};x,0)+
\frac{-11 + 2 x + 12 \epsilon - 2 x \epsilon}{(-1 + x)^2 x}\hat F(\{0,0,0\};x,0)+\\&
\frac{ - 12 (-1 + \epsilon)}{(-1 + x)^2 x}\hat F(\{-1,0,0\};x,0)+
\frac {-  36 (-1 + \epsilon)}{(-1 + x)^2 x}\hat F(\{0,-1,0\};x,0).
\end{split}
\end{align}
With the boundary condition at $x=1/2$ given in Eq.~\eqref{eq:check1}, solving the above DE can also obtain $\hat F(\{0,0,1\};x,0)$  at any value of $x$.

\subsection{VRR}\label{sec:VRRR}

\begin{figure}[htb!]
\begin{center}
\includegraphics[width=0.6\textwidth]{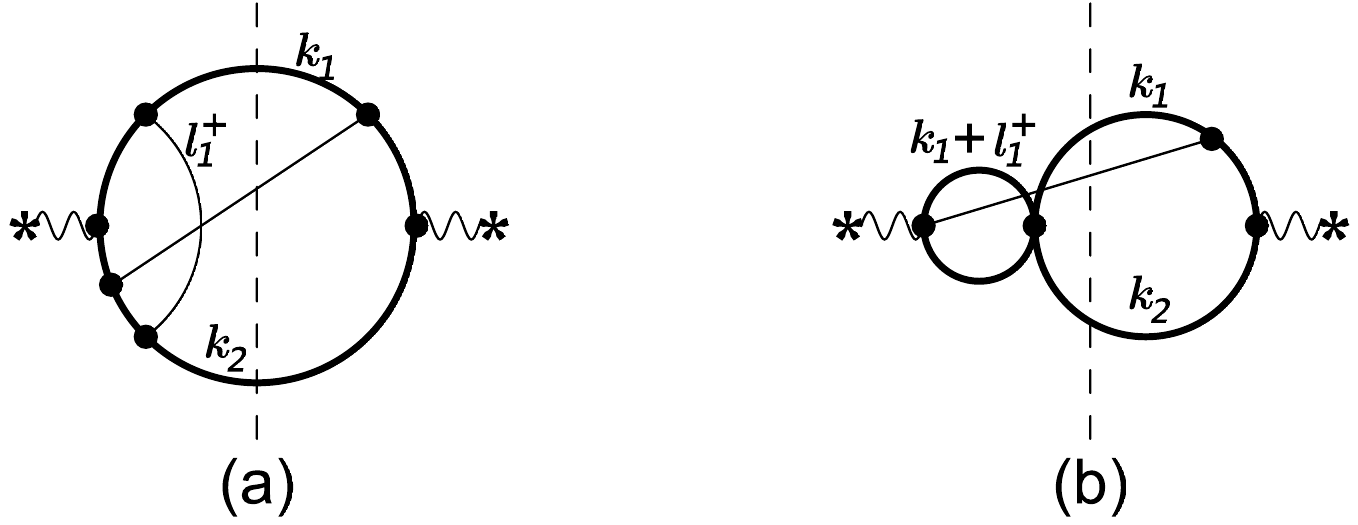}
\caption{\label{fig:VRRR}Representative Feynman diagrams  in VRR, where (a) defines the most complicated family and (b) defines a sub-family of (a). Here thick curves represent top quark, thin curves represent massless particle, and vertical dashed lines represent final state cut.}
\end{center}
\end{figure}

For  VRR, inverse propagators can be written as
\begin{align}
&\mathcal{D}^{\mathrm{c}}_1=k_1^2-m_t^2,
\mathcal{D}^{\mathrm{c}}_2=k_2^2-m_t^2,
\mathcal{D}^{\mathrm{c}}_3=(Q-k_1-k_2)^2;\,\nonumber\\
&\mathcal{D}^{\dt}_1=(Q-k_2)^2 - m_t^2,
\mathcal{D}^{\dt}_2=( Q-k_1 )^2- m_t^2;\,\nonumber\\
&\mathcal{D}^+_1= (k_1 + l^+_1)^2 - m_t^2,
\mathcal{D}^+_2=  (k_2 - l^+_1 )^2-m_t^2 ,
\mathcal{D}^+_3= {l^+_1}^2,
\mathcal{D}^+_4=(Q-k_1-k_2 + l^+_1)^2,\,\nonumber\\&
\mathcal{D}^+_5=(Q-k_2 + l^+_1)^2 - m_t^2,
\mathcal{D}^+_6= ( Q-k_1 - l^+_1 )^2-m_t^2 ,
\end{align}
where  $Q^2=s$. Then the 1-loop $1\to3$ phase-space integrals can be expressed as
\begin{align}
\hat{F}(\vec{\nu};x,y)=s^{3-\frac{3}{2}D+\sum \nu^{\mathrm{t}}_{\alpha}+\sum \nu^+_{\alpha}} \int\text{dPS}_3\prod_{\alpha=1}^{2}\frac{1}{(\mathcal{D}^{\mathrm{t}}_{\alpha})^{\nu^{\mathrm{t}}_{\alpha}}}
\int\frac{\ud^Dl^+_1}{(2\pi)^D}\prod_{\alpha=1}^{6}\frac{1}{(\mathcal{D}^+_{\alpha}+\ui\eta)^{\nu^+_{\alpha}}}.
\end{align}
We find that there are 71 MIs for finite $\eta$ (or $y$) and the number is reduced to 27 when $\eta\to 0^+$.
These MIs can be classified into 2 families
\begin{align}
\{ 1, 0, 1, 1, 1, 0, 0, 1\}~~~\text{and}~~~\{ 1, 0, 0, 1, 0, 1, 1, 1\}.
\end{align}
The most complicated family is the first family, which can also be characterized by Feynman diagram Fig.\ref{fig:VRRR} (a).
Using our method, the result of the most complicated MI with, e.g., $x=1/2$ is
\begin{normalsize}
\begin{align}
\begin{autobreak}
\MoveEqLeft
\hat{F}(\{ 2, 0, 1, 1, 1, 0, 0, 1\};\frac{1}{2},0)=\hspace{10cm}
( 4.35941187166437229714484148598\times10^{-6} -
  4.87955595721663859057448350469\times10^{-6} \ui)\epsilon^{-2} 
  +( 0.000290878052291807102955726741096 +
  0.000037713113716691251718223752983 \ui)\epsilon^{-1} 
  +(0.00232637225490549068317799097260 +
   0.00161913500108049877728334443544 \ui) 
   +(0.0064623807207395294287699841138 +
    0.0143681169585383071360453670293 \ui) \epsilon 
    -(0.0244064366687345660260807481505 -
    0.0681410637818588674321247511745 \ui) \epsilon^2
     -(0.324062211265101500673336745067 -
    0.176280502471673558710833935083 \ui) \epsilon^3
     -(1.84537524457279855776436524767 -
    0.05103632828295095891823713650 \ui) \epsilon^4
     -(7.60322033887241114962526507189 +
    2.00523602305209729225222600674 \ui) \epsilon^5
     -(27.2739377963678526029153832666 +
    12.0357468364668313721542998217 \ui) \epsilon^6+\cdots.
\end{autobreak}
\end{align}
\end{normalsize}

In the following, let us take a sub-family $\{1,0,1,1,0,0,0,0\}$ shown in Fig.\ref{fig:VRRR} (b) as an example to illustrate the calculation procedure of MIs.
For brevity, we define MIs for this family as:
\begin{align}
\hat F(\{\nu^{\dt}_1,\nu^+_1,\nu^+_2\};x,y)=s^{3-\frac{3}{2}D+\nu^{\dt}_1+\nu^+_1+\nu^+_2}
\int\text{dPS}_3\frac{1}{(\mathcal{D}^{\mathrm{t}}_1)^{\nu^{\mathrm{t}}_1}}
\int\frac{\ud^Dl^+_1}{(2\pi)^D}\frac{1}{(\mathcal{D}^+_{1}+\ui\eta)^{\nu^+_1}(\mathcal{D}^+_{2}+\ui\eta)^{\nu^+_2}}.
\end{align}
This family contains 7 MIs for finite $\eta$ 
\begin{align}
\begin{split}
&\left\{
\hat F(\{0,0,1\};x,y),\hat F(\{-1,0,1\};x,y),\hat F(\{0,1,1\};x,y),\hat F(\{-1,1,1\};x,y),\hat F(\{0,1,2\};x,y),\right.\\&\left.
\hat F(\{1,1,1\};x,y),\hat F(\{1,1,2\};x,y)
\right\},
\end{split}
\end{align}
and 6 MIs as $\eta\to0^+$
\begin{align}
\begin{split}
&\left\{\hat F(\{0,0,1\};x,0),\hat F(\{-1,0,1\};x,0),\hat F(\{0,1,1\};x,0),\hat F(\{0,1,2\};x,0),
\hat F(\{1,1,1\};x,0),\hat F(\{1,1,2\};x,0)\right\}.
\end{split}
\end{align}

To calculate the last 2 physical MIs, we set up DEs for corresponding MIs with $x=1/2$ and finite $\eta$, which gives
\begin{align}\label{eq:dey2}
\begin{split}
\frac{\partial}{\partial{y}}\hat F(\{1,1,1\};\frac{1}{2},y)=& -2 \ui\hat F(\{1,1,2\};\frac{1}{2},y),\\
\frac{\partial}{\partial{y}}\hat F(\{1,1,2\};\frac{1}{2},y)=&
\frac{  16 \ui (1 - 2 \epsilon) ( 3 \epsilon-1)}{(-\ui + 8 y) (\ui +  8 y)}\hat F(\{1,1,1\};\frac{1}{2},y)+
\frac{ 4 (\ui + 8 y - 4 \ui \epsilon - 64 y \epsilon)}{(-\ui + 8 y) (\ui +   8 y)}\hat F(\{1,1,2\};\frac{1}{2},y)\\&
-\frac{64 (5 - 11 \epsilon + 6 \epsilon^2)}{ y (-\ui + 8 y) (\ui + 8 y)}\hat F(\{0,0,1\};\frac{1}{2},y)+
\frac{768 (-1 + \epsilon)^2}{ y (-\ui + 8 y) (\ui + 8 y)}\hat F(\{-1,0,1\};\frac{1}{2},y)\\&
+\frac{ 8 (1 - 2 \epsilon) (5 - 4 \ui y -6 \epsilon + 8 \ui y \epsilon)}{ y (-\ui + 8 y) (\ui + 8 y)}\hat F(\{0,1,1\};\frac{1}{2},y)
+\frac{  32 (1 - 2 \epsilon) (-3 + 4 \epsilon)}{  y (-\ui + 8 y) (\ui + 8 y)}\hat F(\{-1,1,1\};\frac{1}{2},y)\\&
-\frac{4 (-\ui + 4 y) (-1 + 2 \epsilon)}{ y (-\ui + 8 y)}\hat F(\{0,1,2\};\frac{1}{2},y).
\end{split}
\end{align}
The boundary condition of these DEs is given by
\begin{align}
\begin{split}
\hat F(\{1,1,1\};\frac{1}{2},y)&\xlongequal{\eta\sim\infty}s^{6-\frac{3}{2}D}
\eta^{\frac{D}{2}-2}\frac{\ui(D-2)}{2}F^{\mathrm{bub}}_{1,1}(D)\left(\int\text{dPS}_3\frac{1}{\mathcal{D}^\dt_1}\right)_{x=1/2},\\
\hat F(\{1,1,2\};\frac{1}{2},y)&\xlongequal{\eta\sim\infty}s^{7-\frac{3}{2}D}
\eta^{\frac{D}{2}-3}\frac{(4 -D) (D-2)}{8}F^{\mathrm{bub}}_{1,1}(D)\left(\int\text{dPS}_3\frac{1}{\mathcal{D}^\dt_1}\right)_{x=1/2},
\end{split}
\end{align}
where
\begin{align}\label{eq:loopbub1}
F^{\mathrm{bub}}_{1,1}(D) \equiv\int\frac{\ud^D{l^+_1}}{(2\pi)^D}\frac{1}{{l^+_1}^2+\ui},
\end{align}
and $\int\text{dPS}_3\frac{1}{\mathcal{D}^\dt_1}$   can again be calculated by the method in Sec.\ref{sec:RRRR} or be obtained from RR in Sec.\ref{sec:RRR}.
Knowing the first 5   MIs, by solving the DEs~\eqref{eq:dey2} with boundary condition we obtain, e.g.,
\begin{normalsize}
\begin{align}\label{eq:etsol}
\begin{autobreak}
\MoveEqLeft
\hat F(\{1,1,2\};\frac{1}{2},0)=
(7.78790446721069262502850093774\times10^{-6} +
   2.91319469772237394135356308348\times10^{-6} \ui) 
   +(0.000130430373015787655604488198861 +
    0.000068404169458201184291920092123 \ui) \epsilon
     +(0.001077434813828191909787186362432 +
    0.000750926876250745472210277589430 \ui) \epsilon^2 
    +(0.00584278150920839062615612508136 +
    0.00527570101382158661589031061691 \ui) \epsilon^3 
    +(0.0233461280012444372334494219123 +
    0.0270859736951617524563966282868 \ui) \epsilon^4 
    +(0.0730918539437148667076104654800 +
    0.1095165249743204252589933869672 \ui) \epsilon^5 
    +(0.185975373883125986488613881520 +
    0.366393770042708443331564801509 \ui) \epsilon^6 
    +(0.393093986188519076512424694564 +
    1.052172170765638116257825410632 \ui) \epsilon^7 
    +(0.69775277299606861048706250047 +
    2.67282546122008383022615104289 \ui) \epsilon^8+\cdots.
\end{autobreak}
\end{align}
\end{normalsize}

Finally, the DEs w.r.t. $x$ is given by
\begin{align}
\begin{split}
&\frac{\partial}{\partial{x}}
\left(
\renewcommand\arraystretch{2.5}
\setlength{\arraycolsep}{1pt}
\begin{array}{c}
\hat F(\{1,1,1\};x,0)\\
\hat F(\{1,1,2\};x,0)
\end{array}
\right)
=
\left(
\renewcommand\arraystretch{2.5}
\setlength{\arraycolsep}{1pt}
\begin{array}{cc}
\frac{ -\epsilon}{ x}&\frac{ 1}{2}
\\
\frac{(1 - 2 \epsilon) ( 3 \epsilon-1)}{(-1 + x) x}&
\frac{x + 4 \epsilon - 10 x \epsilon}{2 (-1 + x) x}
\end{array}
\right)
\left(
\renewcommand\arraystretch{2.5}
\setlength{\arraycolsep}{1pt}
\begin{array}{c}
\hat F(\{1,1,1\};x,0)\\
\hat F(\{1,1,2\};x,0)
\end{array}
\right)+\\
&\left(
\renewcommand\arraystretch{2.5}
\setlength{\arraycolsep}{1pt}
\begin{array}{cccc}
 \frac{ 6 (-1 + \epsilon) (-4 - x + 4 \epsilon + 2 x \epsilon)}{(-1 + x)^2 x^2 (-1 + 2 \epsilon)}&
 \frac{ - 72 (-1 + \epsilon)^2}{(-1 + x)^2 x^2 (-1 + 2 \epsilon)}&
 \frac{ -  2 (-1 + 2 \epsilon)}{(-1 + x) x}&\frac{ -2}{-1 + x}
  \\
   \frac{- 4 (4 + 3 x - 8 \epsilon - 8 x \epsilon + 4 \epsilon^2 +  5 x \epsilon^2)}{(-1 + x)^2 x^3}&
   \frac{ 48 (-1 + \epsilon)^2}{(-1 + x)^2 x^3}&
   \frac{ 2 (1-2\epsilon)^2}{(-1 + x)^2 x}&
   \frac{  2 (-1 + 2 \epsilon)}{(-1 +   x)^2}
\end{array}
\right)
\left(
\renewcommand\arraystretch{1.5}
\setlength{\arraycolsep}{1pt}
\begin{array}{c}
\hat F(\{0,0,1\};x,0)\\
\hat F(\{-1,0,1\};x,0)\\
\hat F(\{0,1,1\};x,0)\\
\hat F(\{0,1,2\};x,0)
\end{array}
\right).
\end{split}
\end{align}
With the boundary condition at $x=1/2$ given in Eq.~\eqref{eq:etsol}, solving the above DEs can also evaluate $\hat F(\{1,1,1\};x,0)$ and $\hat F(\{1,1,2\};x,0)$ at any value of $x$.

\subsection{VVR and VRV}\label{sec:VVRR}
\begin{figure}[htb!]
\begin{center}
\includegraphics[width=0.6\textwidth]{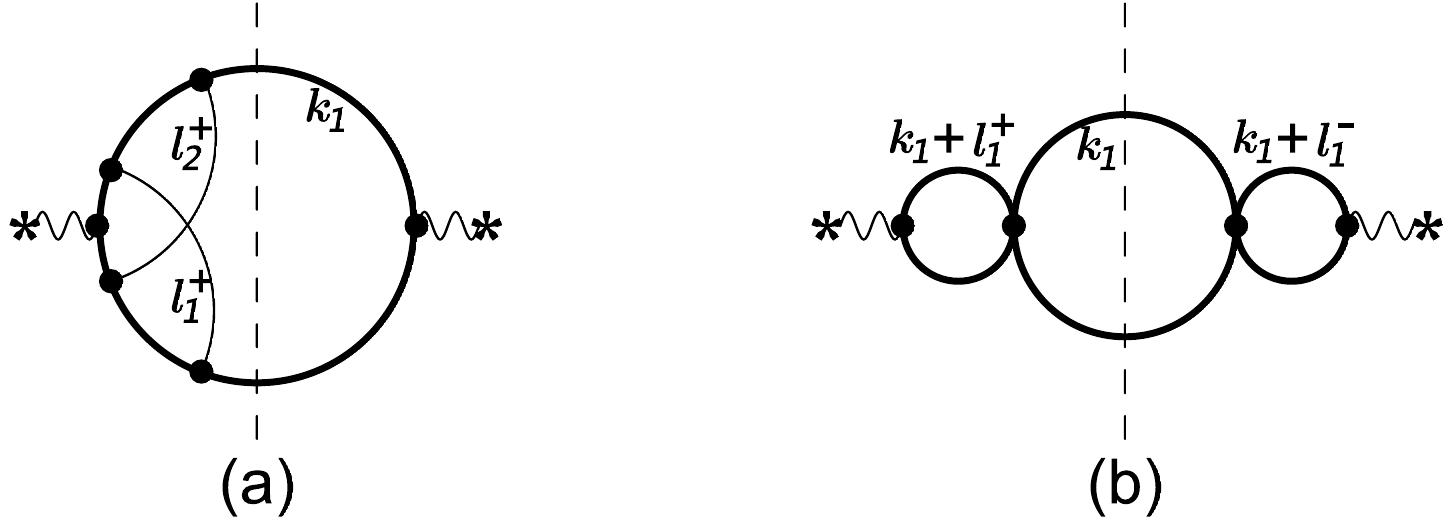}
\caption{\label{fig:VVRRVRRV} 
Representative Feynman diagrams  in VVR and VRV, where (a) defines the most complicated family for VVR and (b) defines the family for VRV. Here thick curves represent top quark, thin curves represent massless particle, and vertical dashed lines represent final state cut.}
\end{center}
\end{figure}

For  VVR,  inverse propagators  can be  written as
\begin{align}
&\mathcal{D}^{\mathrm{c}}_1=k_1^2-m_t^2,
\mathcal{D}^{\mathrm{c}}_2=(Q-k_1)^2-m_t^2;\,\nonumber\\
&\mathcal{D}^+_1= (k_1 + l^+_1 + l^+_2)^2 - m_t^2,
\mathcal{D}^+_2=(k_1 + l^+_2)^2 -  m_t^2,
\mathcal{D}^+_3=(k_1 + l^+_1)^2 - m_t^2,
\mathcal{D}^+_4=  (-k_1 - l^+_1 + Q)^2-m_t^2 ,\,\nonumber\\&
\mathcal{D}^+_5= {l^+_1}^2,
\mathcal{D}^+_6= {l^+_2}^2,
\mathcal{D}^+_7=  (-k_1 - l^+_2 +   Q)^2-m_t^2 ,
\mathcal{D}^+_8=(-k_1 - l^+_1 - l^+_2 + Q)^2-m_t^2 ,
\mathcal{D}^+_9=(l^+_1 + l^+_2)^2 .
\end{align}
Then phase-space integrals  can be expressed as
\begin{align}
\hat F(\vec{\nu};x,y)=s^{2-\frac{3}{2}D+ \sum \nu^+_{\alpha}}
\int\text{dPS}_2
\int\frac{\ud^Dl^+_1\ud^Dl^+_2}{(2\pi)^{2D}}\prod_{\alpha=1}^{9}\frac{1}{(\mathcal{D}^+_{\alpha}+\ui\eta)^{\nu^+_{\alpha}}}.
\end{align}
We  find that there are 53 MIs for finite $\eta$ and the number is reduced to 21 when $\eta\to 0^+$. These MIs can be classified
into 3 families
\begin{align}
\{ 1, 1, 0, 1, 1, 1, 0, 1, 0\} ~~~\text{and}~~~\{ 1, 1, 1, 1, 0, 1, 0, 0,1\}~~~\text{and}~~~\{1, 1, 1, 0, 0, 1, 1, 1, 0\}.
\end{align}
The most complicated family is the first family, which can also be characterized by Feynman diagram Fig.\ref{fig:VVRRVRRV} (a). We use the same method (note, $F^{\mathrm{bub}}$ up to 2-loop is needed to calculate boundary conditions) in  VRR to calculate MIs for these families.

For  VRV,  inverse propagators  can be  written as
\begin{align}
&\mathcal{D}^{\mathrm{c}}_1=k_1^2-m_t^2,
\mathcal{D}^{\mathrm{c}}_2=(Q-k_1)^2-m_t^2;\,\nonumber\\
&\mathcal{D}^+_1=(k_1 + l^+_1)^2 - m_t^2,
\mathcal{D}^+_2= (k_1 + l^+_1 - Q)^2 -m_t^2 ,
\mathcal{D}^+_3= {l^+_1}^2;\,\nonumber\\
&\mathcal{D}^-_1= (k_1 + l^-_1)^2-m_t^2 ,
\mathcal{D}^-_2= (k_1  + l^-_1- Q)^2-m_t^2 ,
\mathcal{D}^-_3= {l^-_1}^2,
\end{align}
in addition to which there is a scalar product $l_1^+\cdot l_1^-$. The obtained family is:
\begin{align}
\{1,1,0,1,1,0,0\},
\end{align}
which can also be characterized by Feynman diagram Fig.\ref{fig:VVRRVRRV} (b).
Because loop integrations in MIs of this case are factorized, their calculation is as simple as one-loop case.

\subsection{Verification of the results}\label{sec:check}

On the one hand, the AMF method can be used to calculate MIs at any given value of $x$. On the other hand, MIs at different values of $x$ can be related by DEs w.r.t. $x$. We have verified that results of MIs for different values of $x$, e.g., $x=1/2$ and $x=2/5$, obtained by the two strategies are consistent with each other. This provides a highly nontrivial self-consistent check because DEs w.r.t. $\eta$ and that w.r.t. $x$ are significantly different.

Our numerical values for  $\hat F(\vec{\nu};0,0)$, i.e. MIs for massless QCD, are in full agreement with the known analytical  results of massless MIs in literature \cite{Gehrmann-DeRidder:2003pne,GehrmannDeRidder:2004tv}.

For RRR sub-process (Sec.\ref{sec:RRRR}), our numerical results for $\hat F(\{0,0,0,0,0,0,1\};x,0)$ and $\hat F(\{0,0,0,1,0,0,1\};x,0)$ show excellent agreement with  the corresponding analytical results in literature \cite{Bernreuther:2011jt} ($T_4$ and $T_5$ in this reference).




\section{Summary and outlook}\label{sec:pssum}

In this paper, we extend the AMF method originally developed for Feynman loop integrals \cite{Liu:2017jxz}  to calculate MIs involving also phase-space integration. As a pedagogical example, we use it to calculate MIs encountered in $e^+e^-\rightarrow \gamma^* \rightarrow t\bar{t}+X$ at NNLO. Our results agree with results obtained by using other methods (our fully results are available as ancillary file in the arXiv version). Although AMF method depends on reduction procedure to decompose all integrals to MIs and to set up DEs of MIs w.r.t. $\eta$, the efficiency of reduction has been significantly improved thanks to the recently proposed search algorithm \cite{Liu:2018dmc,Guan:2019bcx}.

It is clear that the AMF method can be used to calculate MIs of any  process, as systematic as the sector decomposition method. However, comparing with the later method, AMF method is much more efficient and can provide very high precision. The high-precision nature makes it possible to obtain analytical results with proper ansatz. For problems with two or more kinematical invariants, where differential equation method works, the AMF method can not only systematically provide as many as needed boundary conditions for DEs w.r.t. kinematical invariants, but also provide highly nontrivial self-consistent check for the obtained results. All the above advantages make the AMF method very useful for perturbative calculation at high orders.

\begin{acknowledgments}
We thank Chen-Yu Wang, K.T. Chao and C. Meng, Xin Guan, Xiao-Bo Jin, Zhi-Feng Liu  for many helpful communications and discussions.
The work is supported in part by the National Natural Science Foundation of China (Grants No. 11875071, No. 11975029) and the High-performance Computing Platform of Peking University.
\end{acknowledgments}

\appendix

\section{Calculation of $F^{\mathrm{cut}}$}\label{sec:pscut}

In this appendix, we calculate MIs of basal phase-space integration with no denominator.  It is important to note that, besides masses presented in cut lines, these MIs only depend on the square of center of mass energy $s\equiv Q^2$, regardless of the configuration of external unintegrated momenta. We use $F_{r,N,n}^{\mathrm{cut}}$ to denote the $n$-th MI for $N$-particle-cut integrals with $m_1=\cdots=m_r=m$ and $m_{r+1}=\cdots=m_N=0$, and we will provide explicit results for $r=0,~1$, and $2$. MIs for general cases can be studied as following. By using optical theorem (see, e.g., Ref.~\cite{Gehrmann-DeRidder:2003pne}), the calculation of basal phase-space integral is translated to the calculation of imaginary part of the corresponding sunrise pure loop integral. The later can be calculated by the AMF method for loop integrals~\cite{Liu:2017jxz}. Furthermore, a one-dimensional-integral representation of all these MIs can be obtained from Ref.~\cite{Groote:2005ay}.

For $r=0$ and $N\geq2$, theres is only one MI, which is given by
\begin{align}\label{eq:massless}
F_{0,N,1}^{\mathrm{cut}}\equiv\int\text{dPS}_N=
\frac{2^{5-4N-2\epsilon+2N\epsilon}\pi^{3-2N-\epsilon+N\epsilon}
\Gamma(1-\epsilon)^N}{\Gamma((N-1)(1-\epsilon))\Gamma(N(1-\epsilon))}s^{N-2+\epsilon-N\epsilon}.
\end{align}
For $r=1$, in general there are two MIs, which are given by ($n=1,2$)
\begin{align}\label{eq:f1n}
\begin{split}
F^{\mathrm{cut}}_{1,N,n}\equiv&\int\text{dPS}_N\left((Q-k_1)^2\right)^{n-1}\\
=&\frac{2^{2(5-n-3N-2\epsilon+2N\epsilon)}\pi^{\frac{7}{2}+N(\epsilon-2)-\epsilon}\Gamma(1-\epsilon)^{N-1}\Gamma(n-3+N+2\epsilon-N\epsilon)}{
\Gamma((N-1)(1-\epsilon))\Gamma(n-\frac{3}{2}+N+\epsilon-N\epsilon)\Gamma((N-2)(1-\epsilon))}\\&
\times s^{n-3+N+\epsilon-N\epsilon}\left(1-\frac{m^2}{s}\right)^{2(n+N+\epsilon-N\epsilon)-5}\\&
\times {}_2F_1\left(n-2+N+\epsilon-N\epsilon,n-3+N+2\epsilon-N\epsilon;2(n-2+N+\epsilon-N\epsilon);1-\frac{m^2}{s}\right),
\end{split}
\end{align}
where  ${}_pF_q$ are generalized hypergeometric functions which can be evaluated by using the public program HypExp \cite{Huber:2007dx}. In the special case of $N=2$, there is only one MI $F^{\mathrm{cut}}_{1,2,1}$.
For $r=2$, in general there are three MIs, which are given by ($n=1,2,3$)
\begin{align}\label{eq:f2n}
\begin{split}
F^{\mathrm{cut}}_{2,N,n}\equiv&\int\text{dPS}_N\left((k_1+k_2)^2\right)^{n-1}\\
=&\frac{2^{5+2N(\epsilon-2)-2\epsilon}\pi^{3+N(\epsilon-2)-\epsilon}\Gamma(1+n-2\epsilon)\Gamma(1-\epsilon)^{N-1}\Gamma(n-\epsilon)}{\Gamma(2-2\epsilon)
\Gamma(n-1+N-N\epsilon)\Gamma(n-2+N+\epsilon-N\epsilon)}
s^{n-3+N+\epsilon-N\epsilon}\\&
~~~\times {}_3F_2\left(\epsilon-\frac{1}{2},2-n-N+N\epsilon,3-n-N-\epsilon+N\epsilon;1-n+\epsilon,2\epsilon-n;\frac{4m^2}{s}\right)\\&
+\frac{2^{4+2N(\epsilon-2)}\pi^{\frac{7}{2}+N(\epsilon-2)-\epsilon}\Gamma(1-\epsilon)^{N-1}\Gamma(\epsilon-n)}{\Gamma(\frac{3}{2}-n)
\Gamma((N-1)(1-\epsilon))\Gamma((N-2)(1-\epsilon))}
s^{n-3+N+\epsilon-N\epsilon}\left(\frac{4m^2}{s}\right)^{n-\epsilon}\\&
~~~\times {}_3F_2\left(n-\frac{1}{2},3-N-2\epsilon+N\epsilon,2-N-\epsilon+N\epsilon;1+n-\epsilon,\epsilon;\frac{4m^2}{s}\right)\\&
+\frac{2^{4+2N(\epsilon-2)}\pi^{\frac{7}{2}+N(\epsilon-2)-\epsilon}\Gamma(1-\epsilon)^{N-2}\Gamma(\epsilon-1)\Gamma(2\epsilon-1-n)}{\Gamma(\frac{1}{2}-n+\epsilon)
\Gamma((N-2)(1-\epsilon))\Gamma((N-3)(1-\epsilon))}
s^{n-3+N+\epsilon-N\epsilon}\left(\frac{4m^2}{s}\right)^{1+n-2\epsilon}\\&
~~~\times {}_3F_2\left(\frac{1}{2}+n-\epsilon,4-N-3\epsilon+N\epsilon,3-N-2\epsilon+N\epsilon;2+n-2\epsilon,2-\epsilon;\frac{4m^2}{s}\right).
\end{split}
\end{align}
In the special case of $N=2$, there are only one MI, which means $F^{\mathrm{cut}}_{2,2,2}$ and $F^{\mathrm{cut}}_{2,2,3}$ can be reduced to $F^{\mathrm{cut}}_{2,2,1}$. In the special case of $N=3$, there are only two MIs, which means $F^{\mathrm{cut}}_{2,3,3}$ can be reduced to $F^{\mathrm{cut}}_{2,3,1}$ and $F^{\mathrm{cut}}_{2,3,2}$. The number of MIs obtained here is consistent with the expectation in Ref.~\cite{Kalmykov:2016lxx}. The results  \eqref{eq:f1n}  and \eqref{eq:f2n} agree with partial results in literature [see Eq.(B.7) in Ref.~\cite{Anastasiou:2013srw} and  Eqs. (3.14-3.16) in Ref.\cite{Bernreuther:2011jt}].

To obtain \eqref{eq:f1n}  and \eqref{eq:f2n}, we decompose the measure of phase-space integration to two parts:
\begin{align}\label{eq:psdec}
\begin{split}
\text{dPS}_N(Q;\{k_i\})\xlongequal[s=Q^2]{s_A=k_A^2}\int_{(\sum_{i\in\Omega}m_i)^2}^{(\sqrt{s}-\sum_{i\notin\Omega}m_i)^2}\frac{\ud s_A}{2\pi}\text{dPS}_{N-\#_{\Omega}+1}(Q;  \left\{k_{i\notin\Omega}\right\}, k_A)\text{dPS}_{\#_{\Omega}}(k_A; \left\{k_{i\in\Omega}\right\}),
\end{split}
\end{align}
where $\Omega$ is a subset of $\{1,2,\ldots,N\}$ and $\#_{\Omega}$ is the number of elements of  $\Omega$. Specifically, the $\Omega$ is chosen to be the set of massless particles for $r=1$, and  is chosen to be the set of massive particles for $r=2$. The phase-space volume of two-particle cut is used:
\begin{align}\label{eq:basmif2}
\Phi_2=\int\text{dPS}_2=\frac{2^{-3 + 2\epsilon}\pi^{-1 + \epsilon}\Gamma(1 - \epsilon)}{\Gamma(2 -2\epsilon)}s^{-\epsilon}\left( 1 - \frac {2 (m_ 1^2 + m_ 2^2)} {s} + \left(\frac {m_ 1^2 - m_2^2}{s}\right)^2\right)^{\frac{1}{2}-\epsilon}.
\end{align}
The following relations (can be found, e.g., from Ref.~\cite{Pru}) are also useful:
\begin{align}\label{eq:tran1}
{}_2F_1\left(a,b;c;z\right)&=(1-z)^{-a}{}_2F_1\left(a,c-b;c;\frac{z}{z-1}\right),\\
{}_2F_1\left(a,1-a;c;z\right)&=(1-z)^{c-1}\left(\sqrt{1-z}+\sqrt{-z}\right)^{2-2a-2c}
{}_2F_1\left(a+c-1,c-\frac{1}{2};2c-1;\frac{4\sqrt{z^2-z}}{(\sqrt{1-z}+\sqrt{-z})^2}\right),
\end{align}
\begin{align}\label{eq:intf1n}
\begin{split}
&\int_{z}^{y}\ud{x}x^{\alpha-1}(y-x)^{c-1}(x-z)^{\beta-1}{}_2F_1\left(a,b;c;1-\frac{x}{y}\right)=\\&
~~~\frac{\Gamma(c)\Gamma(c-a-b)\Gamma(\beta)\Gamma(1-\alpha-\beta)}{\Gamma(c-a)\Gamma(c-b)\Gamma(1-\alpha)}y^{c-1}z^{\alpha+\beta-1}\\&
~~~\times {}_3F_2\left(a-c+1,b-c+1,\alpha;a+b-c+1,\alpha+\beta;\frac{z}{y}\right)\\&+
\frac{\Gamma(c)\Gamma(a+b-c)\Gamma(\beta)\Gamma(a+b-c-\alpha-\beta+1)}{\Gamma(a)\Gamma(b)\Gamma(a+b-c-\alpha+1)}y^{a+b-1}z^{c-a-b+\alpha+\beta-1}\\&
~~~\times {}_3F_2\left(1-a,1-b,c-a-b+\alpha;c-a-b+1,c-a-b+\alpha+\beta;\frac{z}{y}\right)\\&+
\frac{\Gamma(c)\Gamma(\alpha+\beta-1)\Gamma(c-a-b+\alpha+\beta-1)}{\Gamma(c-a+\alpha+\beta-1)\Gamma(c-b+\alpha+\beta-1)}y^{c+\alpha+\beta-2}\\&
~~~\times {}_3F_2\left(1-\beta,a-c-\alpha-\beta+2,b-c-\alpha-\beta+2;2-\alpha-\beta,a+b-c-\alpha-\beta+2;\frac{z}{y}\right).
\end{split}
\end{align}

\providecommand{\href}[2]{#2}\begingroup\raggedright\endgroup

\end{document}